\documentclass[prd,reprint,nofootinbib,superscriptaddress]{revtex4-2}
\usepackage{bm}
\usepackage{amsmath}
\usepackage{graphicx}
\usepackage{float}
\usepackage{amssymb}
\usepackage{hyperref}
\usepackage{CJK}
\setcitestyle{super}
\hypersetup{
	colorlinks=true,
	linkcolor=red,
	citecolor=blue,
}

\begin{document}
\begin{CJK*}{UTF8}{gbsn}

\title{Concepts and status of Chinese space gravitational wave detection projects}

\author{Yungui Gong
\href{https://orcid.org/0000-0001-5065-2259}{\includegraphics[scale=0.4]{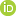}}}
\email{yggong@hust.edu.cn}
\affiliation{School of Physics, Huazhong University of Science and Technology, Wuhan, China}

\author{Jun Luo}
\affiliation{MOE Key Laboratory of TianQin Mission, TianQin Research Center for Gravitational Physics \& School of Physics and Astronomy, Frontiers Science Center for TianQin, CNSA Research Center for Gravitational Waves, Sun Yat-sen University, Zhuhai, China}
\affiliation{Center for Gravitational Experiments, School of Physics, MOE Key Laboratory of Fundamental Physical Quantities Measurement \& Hubei Key Laboratory of Gravitation and Quantum
Physics, PGMF, Huazhong University of Science and Technology, Wuhan, China}

\author{Bin Wang
\href{https://orcid.org/0000-0003-3419-9153}{\includegraphics[scale=0.4]{orcid.png}}}
\email{wang\_b@sjtu.edu.cn}
\affiliation{Shanghai Frontier Science Research Center for Gravitational Wave Detection, School of Aeronautics and Astronautics, Shanghai Jiao Tong University, Shanghai, China}
\affiliation{Center for Gravitation and Cosmology, Yangzhou University, Yangzhou, China.}

\begin{abstract}
Gravitational wave (GW) detection in space probes GW spectrum that is inaccessible from the Earth. In addition to LISA project led by European Space Agency,
and the DECIGO detector proposed by the Japan Aerospace Exploration Agency,
two Chinese space-based GW observatories--
TianQin and Taiji--are planned to be launched in the 2030s.
TianQin has a unique concept in its design with a geocentric orbit.
Taiji's design is similar to LISA,
but is more ambitious with longer arm distance.
Both facilities are complementary to LISA,
considering that TianQin is sensitive to higher frequencies and Taiji probes similar frequencies but with higher sensitivity.
In this Perspective we explain the concepts for both facilities
and introduce the development milestones of TianQin and Taiji projects  in testing extraordinary technologies to pave the way for future space-based GW detections.
Considering that LISA, TianQin and Taiji have similar scientific goals,
all are scheduled to be launched around the 2030s and will operate concurrently,
we discuss possible collaborations among them to improve GW source localization and characterization.
\end{abstract}


\maketitle

\section{Introduction}

After the first direct detection of gravitational wave (GW) in September 2015 \cite{Abbott:2016blz,TheLIGOScientific:2016agk},
there have been 50 reported GW detections by the ground-based Laser Interferometer Gravitational-Wave Observatory (LIGO) and Virgo observatory \cite{LIGOScientific:2018mvr,Abbott:2020niy}.
These GW events were identified as merges of binary black holes (BBH)  \cite{Abbott:2016blz,TheLIGOScientific:2016agk,Abbott:2016nmj,Abbott:2017vtc,Abbott:2017oio,Abbott:2017gyy,LIGOScientific:2020stg,Abbott:2020tfl}, binary neutron stars  \cite{TheLIGOScientific:2017qsa,Abbott:2020uma}, BH-neutron star mergers or the coalescence of a BH with a mystery compact object \cite{Abbott:2020khf}.

For ground-based GW detectors, their high frequency range (10-1,000Hz) implies that even stellar-mass binary mergers emit GWs for only short durations.
Multiple interferometers are needed to confidently detect GWs and eliminate false alarm events.
Using the time delay between multiple  detectors in different locations allows the sky location of the GW source to be pinpointed using triangulation.
The large distance between the advanced Virgo and the advanced LIGO detectors
helps them achieve different orientations due to the Earth's curvature and because their interferometer arms are not parallel to each other it is expected that the polarization of GWs can be extracted from LIGO-Virgo network data \cite{Abbott:2017oio}.
A global network of ground-based GW detectors will play a key role in exploring the population of BHs and neutron stars with GWs in the frequency band around 100 Hz.
In addition to the advanced LIGO \cite{Harry:2010zz,TheLIGOScientific:2014jea} and Virgo \cite{TheVirgo:2014hva} detectors,
the Kamioka Gravitational Wave Detector (KAGRA) \cite{Somiya:2011np,Aso:2013eba} has joined the network of ground-based detectors.
In the near future, LIGO India will also join the network.

Ground-based detectors  are not sensitive to GWs below 1Hz because of terrestrial gravity gradient noise.
A space-based detector is free from such noise and can be made very large,
thereby expanding the frequency range downwards to $10^{-4}$ Hz,
where  exciting GW sources are waiting to be explored.
Space-based observations can probe dynamic coalescing systems,
characterize how they grow, pair up and merge,
and estimate the space density and environments of massive BHs (MBHs).
Covering different GW frequency ranges,
space-based GW observations will provide a unique tool for testing gravity.
Some highly relativistic events,
such as BBH coalescences with masses below $10^5$ solar masses ($M_\odot$),
last a year or longer.
This  long duration allows a single space-based detector to provide directional information as it orbits in the sky during the observation.
The equilateral triangle design of space-based GW interferometers
can provide independent information on the GW polarization.
Furthermore, the improved sensitivities of space-based detectors,
afforded due to the increased interferometer arm length,
will enable improved searches for the cosmological GW background.
Note that if the arm length is comparable or even longer than the wavelength of GWs, then the sensitivity  deteriorates.

\begin{figure}[htbp]
	\centering
	\includegraphics[width=0.5\textwidth]{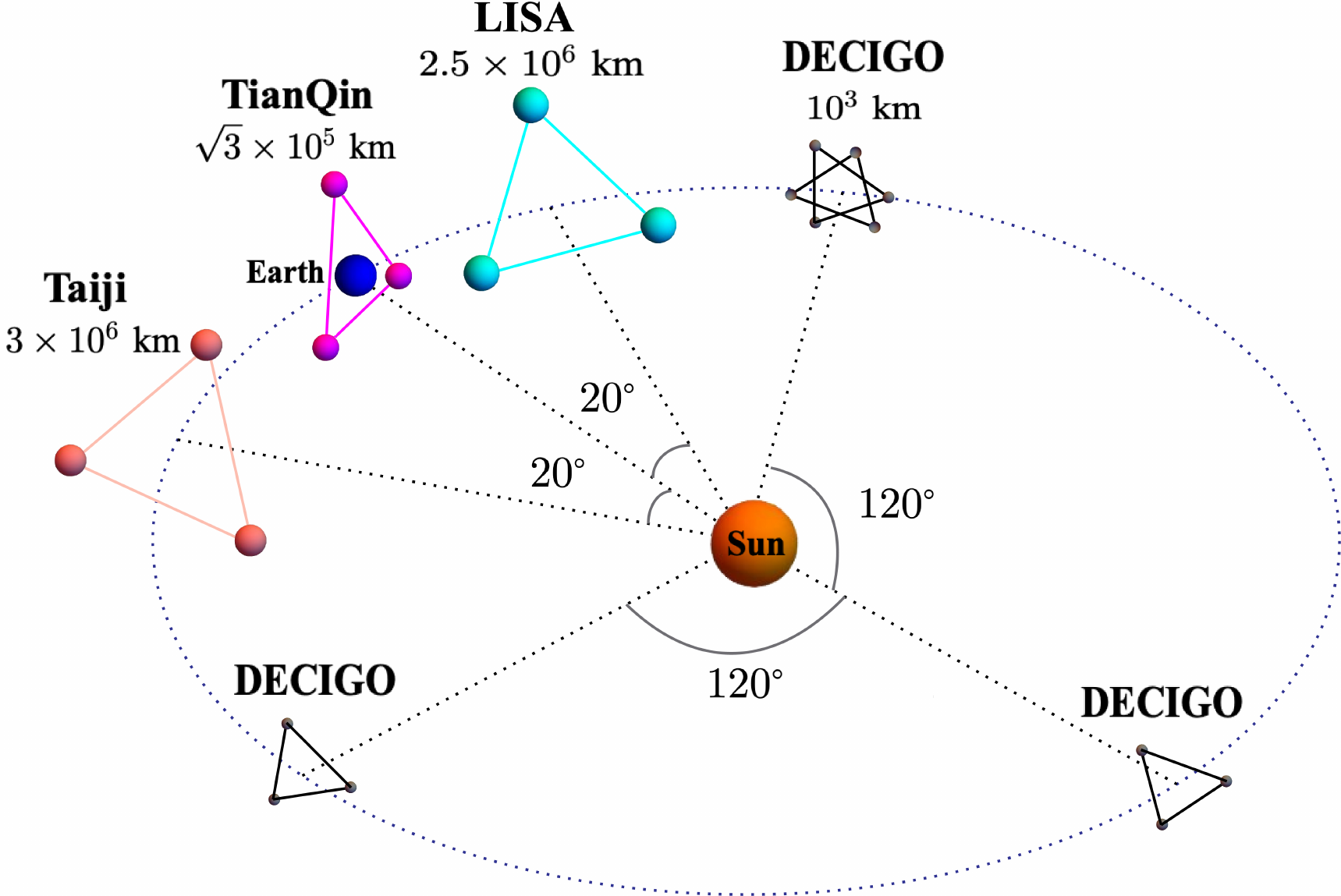}
	\caption{{\bf Schematic of space-based GW detector constellations.}
	The detector plane of TianQin points to the calibration source RX J0806.3+1527 and DECIGO has four clusters.}
\label{tqtjfig}
\end{figure}

Proposed projects for space-based GW observatories include the Laser Interferometer Space Antenna (LISA) \cite{Danzmann:1997hm,Audley:2017drz} led by European Space Agency (ESA),
the Deci-hertz Interferometer Gravitational Wave Observatory (DECIGO) \cite{Seto:2001qf,Kawamura:2006up,Kawamura:2011zz} led by Japan Aerospace Exploration Agency (JAXA) and two Chinese detectors, TianQin \cite{Luo:2015ght} and Taiji \cite{Hu:2017mde}.
LISA consists of three identical spacecraft separated by 2.5 million km that form an equilateral triangle.
The center of the triangle formation is in the ecliptic plane,
1au from the Sun and $20^\circ$ behind the Earth (Fig. \ref{tqtjfig}).
LISA will measure GWs in the band from $10^{-4}$Hz to $10^{-1}$Hz (Fig. \ref{noise}).
To demonstrate the technology capability,
LISA Pathfinder was launched on 3 December 2015.
The mission showed that the residual noise of the interferometer onboard is $32.0^{+2.4}_{-1.7}$ fm Hz$^{-1/2}$,
the differential acceleration noise is $1.74\pm 0.05$ fm s$^{-2}$ Hz$^{-1/2}$ above 2 mHz and $(6\pm 1)\times 10$ fm s$^{-2}$ Hz$^{-1/2}$ at 20 $\mu$Hz,
and the performance of diagnostic subsystem achieves 10 $\mu$K Hz$^{-1/2}$ in the mission band, 1 mHz $<f<$ 30 mHz (refs. \cite{Armano:2016bkm,Armano:2017oco,Armano:2018kix,Armano:2019dxs,Armano:2019ekt,Armano:2021cwl}). The performance of LISA Pathfinder demonstrates the progress of the LISA project.
The LISA observatory is expected to launch around 2035.

DECIGO is the planned Japanese space-based GW antenna mission \cite{Seto:2001qf,Kawamura:2006up,Kawamura:2011zz} that aims to observe GWs in the frequency range from 0.1 Hz to 10 Hz,
which bridges the frequency gap between LISA and terrestrial detectors,
using an arm length of 1,000 km (Fig. \ref{noise}).
The DECIGO observatory consists of four clusters of spacecraft,
each comprising three spacecraft forming an equilateral triangle
and three Fabry-Perot Michelson interferometers.
Two clusters are placed at the same place in a heliocentric Earth-trail orbit,
while the other two
clusters are distributed around the Sun \cite{Kawamura:2020pcg} (Fig. \ref{tqtjfig}).
DECIGO Pathfinder, B-DECIGO, is expected to launch in the 2030s \cite{Kawamura:2020pcg}.

TianQin is a space-based GW detector
developed by a Chinese team at Sun Yat-Sen University \cite{Luo:2015ght}
with some unique features in its project concept.
Similar to the Orbiting Medium Explorer for Gravitational Astronomy ( OMEGA), Geostationary Antenna for Disturbance-Free Laser Interferometry (GADFLI)
 and Geosynchronous Laser Interferometer Space Antenna (gLISA) \cite{Omega,McWilliams:2011ra,Tinto:2014eua},
TianQin will be placed in a geocentric orbit,
which is easily reachable for the implementation and operation  (Fig. \ref{tqtjfig}).
The arm of the equilateral triangle is around $10^5$km,
and the frequency sensitivity band of the detector
overlaps with that of LISA near $10^{-4}$Hz and with that of DECIGO near 0.1 Hz (Fig. \ref{noise}).
Because TianQin's arm length is shorter and its sensitivity in the higher frequency regime is better than LISA and Taiji,
TianQin bridges the frequency gap between LISA and DECIGO
and might perform better than LISA in the high-frequency regime \cite{Zhang:2020hyx,Zhang:2020drf},
that is relevant to the search for intermediate-mass BHs.
The launch of TianQin is expected to be around 2035.

There is another Chinese project of space-based GW detector, Taiji,
which was proposed by scientists at the Chinese Academy of Sciences \cite{Hu:2017mde}.
Similar to LISA, Taiji is in a heliocentric orbit ahead of the Earth by about 18--20$^\circ$ (Fig. \ref{tqtjfig}),
as a compromise between orbit injection costs and Earth-lunar system disturbance.
The spacecraft are farther apart than LISA with a separation distance of 3 million km,
giving the detector access to frequencies covering the range of $10^{-4}$Hz to 0.1Hz with higher sensitivity around $0.01-0.1$ Hz than LISA (Fig. \ref{noise}).
If both launch according to plan,
Taiji and LISA will have a few years of overlapping observing time,
leading to some key capabilities that will be discussed below.
Taiji is expected to launch around 2033.

The Chinese space GW detection projects will mostly be supported by the Chinese government,
including the Ministry of Science and Technology,
the Chinese Academy of Sciences,
the National Natural Science Foundation,
the Ministry of Education,
the China National Space Administration and local government agencies.
For Taiji, the estimated project cost is around 14 billion yuan (US\$2 billion),
roughly the same as the budget for LISA.
The total cost of TianQin is about 2 billion yuan (US\$0.28 billion).
Note that all costs given here are only estimations.

With all three space-based detectors launching in the 2030s,
a period of concurrent observations may be possible,
leading to a true multi-band approach to GW astronomy from space \cite{AmaroSeoane:2009ui,Key:2008tt,Sesana:2016ljz}.
Combining these GW observatories with X-ray, optical and radio instruments such as the Athena mission \cite{Nandra:2013jka,McGee:2018qwb},
the Global Astrometric Interferometer for Astrophysics (GAIA) \cite{GAIA},
the Large Synoptic Sky Survey (LSST) \cite{Abell:2009aa},
the Square Kilometer Array (SKA) \cite{Smits:2008cf},
the European Extremely Large Telescope (EELT) \cite{Evans:2013tua} and so on would provide
multi-messenger observations of the Universe.
Overlap of science targets in different space-based GW observatories will require a networked analysis of data from different space-based GW observatories.
The Taiji group has outlined the possibility of a direct collaboration with LISA \cite{Ruan:2020smc,Wang:2020fwa,Wang:2020dkc}.
The properties of the LISA and TianQin instruments have also been compared and pitted against each other \cite{Zhang:2020hyx,Zhang:2020drf}
and the potential for exploring the unseen universe
is far greater via cooperation than competition:
the space-based detection of GWs should by all means be international.

In addition to millihertz--kilohertz band,
pulsar timing arrays (PTAs) are used to detect GWs
produced by binary supermassive black holes (SMBHs) in the nanohertz frequency band.
These PTAs include the European Pulsar Timing Array (EPTA) \cite{Kramer:2013kea},
the North American Nanohertz Observatory for Gravitational Waves (NANOGrav) \cite{Jenet:2009hk},
the Parkes Pulsar Timing Array (PPTA) \cite{Hobbs:2008yn},
the Five-hundred-meter Aperture Spherical Telescope (FAST) \cite{Hobbs:2014tqa},
and the SKA \cite{Smits:2008cf}.
The EPTA, PPTA and NANOGrav constitute the International Pulsar Timing Array project (IPTA) \cite {Hobbs:2009yy} (Fig. \ref{noise}).
As the stochastic cosmic GW background,
the primordial GWs (the quantum fluctuations of tensor perturbations)
expected from an inflationary scenario \cite{Guth:1980zm,Starobinsky:1980te,Albrecht:1982wi,Linde:1983gd,Sato:1980yn} could also be explored by
the detection of B-mode polarization in the cosmic microwave background radiation,
although it has not been detected yet \cite{Akrami:2018vks,Ade:2018gkx}.
Projects aiming to detect the B-mode polarization include BICEP3 at the South Pole \cite{Ahmed:2014ixy}, SPTpol on the South Pole Telescope (SPT) \cite{Bleem:2012gio},
POLARBEAR-2 \cite{Suzuki:2015zzg},
the Ali project \cite{Cai:2016hqj} at Tibet, China,
the Probe of Inflation and Cosmic Origins (PICO) \cite{Sutin:2018onu},
the LiteBIRD Satellite Mission \cite{Matsumura:2013aja},
the CMB-S4 \cite{Abazajian:2020dmr} and son on.

In the following, we will detail the concept,
roadmap, current status and future development of TianQin and Taiji.
We will also briefly introduce the organizations and collaborations involved in the two projects.
Finally, we will discuss the effectiveness of future collaborations among global space-based GW observatories in accurately locating GW sources.

\section{TianQin project}

TianQin (天琴), meaning `harp in the sky' in Chinese,
is a space-based observatory waiting to be `plucked' by GWs.
This project was first proposed in 2014 \cite{Luo:2015ght},
when GWs were not yet detectable.
The initial goal of the project was to detect a GW signal,
thus the normal vector of TianQin's detector plane points towards the calibration source,
RX J0806.3+1527 (a particular pair of orbiting white dwarf stars, called HM Cancri \cite{Israel:2002gq,Barros:2004er,Roelofs:2010uv,Esposito:2013vja,Kupfer:2018jee}),
the most accessible GW source in the millihertz band.
Different from other space-based GW detectors,
such as LISA and DECIGO,
which aim to detect GWs from unknown sources,
the role of the calibration source will help to determine the sensitivity of the TianQin detector so that it  optimizes the detection of GWs from the
calibration source with known properties.

The three spacecraft of TianQin orbit the Earth with a radius of $\sim 10^5$ km and rotate around the Sun together with the Earth with
an arm length of $\sqrt{3}\times 10^5$ km.
The benefits of choosing a geocentric orbit include reducing the launch cost,
shortening the signal transfer duration,
simplifying telecommunications and more easily guiding satellites through global navigation,
laser ranging and so on.
However, the geocentric orbital motion means that we will have to cope with
a few issues that will not affect LISA and Taiji:
(1) the thermal stability of the spacecraft,
which can be caused by the variation in the sunlight direction relative to the orbital plane or eclipses due to the Earth or the Moon temporarily blocking the sunlight;
(2) the constellation stability relating to the distortion in the equilateral triangle induced by the gravitational disturbances,
especially from the nearby Earth-Moon system;
(3) the problem of a steady power supply because of the eclipses of the Earth and the Moon when the constellation passes through their shadows.

\begin{figure}
	\centering
	\includegraphics[width=0.45\textwidth]{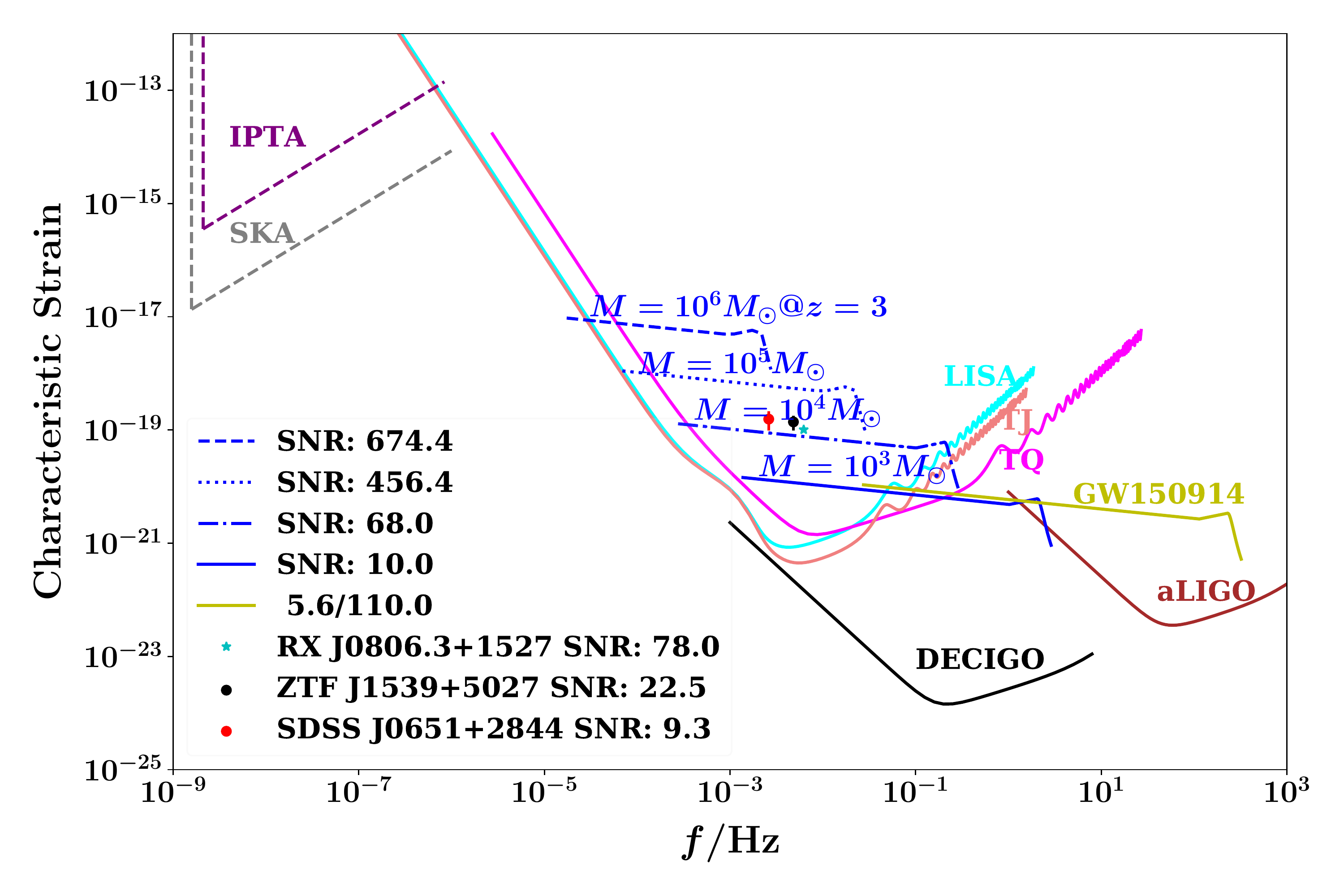}
	\caption{{\bf Noise curves along with various sources.}
We use fitted sensitivity curve for DECIGO \cite{Yagi:2011wg}
and the noise curves for PTAs from ref. \cite{Moore:2014eua}.
The blue lines denote the signals from BBHs with equal mass at $z=3$.
All of the signals start from 1 yr before the merger.
The SNRs of these sources in TianQin were calculated with 1 yr observation time.
For the GW150914-like source, the SNRs in TianQin and advanced LIGO (aLIGO) are about 5.6 and 110, respectively. TQ, TianQin. TJ, Taiji.}
\label{noise}
\end{figure}

To avoid these problems or mitigate their impact,
very careful orbit optimizations and controls have been studied in a series of papers \cite{tqijmpd19,tqijmpd20,Zhang:2020paq,Ye:2020tze}.
It was confirmed that choosing
high Earth orbits can free TianQin from the disturbing effect of the nearby gravitational field of the Earth-Moon system,
because the Newtonian gravity-field environment at $10^5$km from the Earth is fairly quiet and the effect of the Earth-Moon system's gravity field dominates
at frequencies below $10^{-4}$Hz and falls off rapidly towards high frequencies  \cite{Zhang:2020paq}.
To reduce the impact of eclipses on the thermal stability and steady power supply
of the spacecraft, two strategies for choosing the orbit were proposed \cite{Ye:2020tze}.
The first selects the initial phase to be $\sim 15^\circ$.
The second one resizes the orbit to 1:8 synodic resonance with the Moon,
which raises TianQin’s preliminary orbital radius of $1\times 10^5$ km
to 100900 km.
With these two strategies, eclipse-free operation can be maintained during the consecutive three months on/three months off observation windows (continuous observation for three months,
then a shift to safe mode for three months before the next observation restarts).
Under this scheme, the total duration of data acquisition will be 2.5 yr throughout a 5-yr mission,
and the stability of the constellation can be ensured.
Choosing a three-month science run to detect GWs emitted from the calibration source
with a signal-to-noise ratio (SNR) of about 10,
the design requirement for the acceleration noise is
$\sqrt{S_a}=10^{-15}\text{m s}^{-2}\ \text{Hz}^{-1/2}$ and
for the displacement noise is $\sqrt{S_x}=1\ \text{pm Hz}^{-1/2}$.
The noise curve is shown in Fig. \ref{noise}.
In Fig.  \ref{noise},
we also show various sources and their SNRs as expected to be measured by TianQin.

In addition to the calibration source,
TianQin is capable of detecting GWs in the frequency range 0.1 mHz$-$0.1 Hz.
Major sources of such GWs include Galactic binaries such as double white dwarf binaries \cite{Huang:2020rjf},
stellar-mass BH binaries \cite{Liu:2020eko},
the inspiral, merger and ringdown of intermediate-mass BH binaries,
MBH binaries \cite{Wang:2019ryf,Shi:2019hqa,Bao:2019kgt}, SMBH binaries \cite{Feng:2019wgq},
extreme-mass-ratio inspirals \cite{Fan:2020zhy}, GWs from early Universe \cite{Gong:2017qlj,Lin:2020goi},
exotic sources such as cosmic strings \cite{Hu:2017yoc,Ellis:2020ena}
and unmodelled sources.
Expected key science drivers for
TianQin are studies of the seeds and growth of BHs,
the no hair theorem, the nature of gravity,
the expansion of the Universe, and even high-energy physics in the early Universe \cite{Hu:2017yoc,Shi:2019hqa,Bao:2019kgt,Mei:2020lrl}.
The unique frequency range covered by TianQin makes it ideally suited to multi-band astronomy.

To realize these scientific goals,
the project roadmap for TianQin was announced in 2015,
called the 0123 plan,
to develop key technologies strategically and systematically  \cite{Mei:2020lrl}:
\begin{itemize}
  \item Step 0: construct a laser ranging station on the ground to explore the
  high precision orbital information for TianQin through a lunar laser ranging experiment.
  \item Step 1: launch a single satellite primarily to test and demonstrate the readiness of the inertial reference technology.
  \item Step 2: launch a pair of satellites to test and demonstrate the readiness of the intersatellite laser interferometry technology.
  \item Step 3: launch three satellites to form the TianQin constellation to detect GWs.
\end{itemize}

Work towards step 0 started in 2016. The TianQin team manufactured a 17 cm single hollow corner-cube retro-reflector and
installed it on the Chang'e-4 relay satellite, QueQiao, which was launched on 21 May 2018 and reached the halo orbit at the L2 Lagrange point.
The TianQin team also constructed a 5,000m$^2$ (1.2 acre) laser rangefinder station on top of Fenghuang mountain in the Zhuhai campus of Sun Yat-Sen University.
The station contains a 1.20 m (1.3 yd) reflector telescope,
a multichannel superconducting single-photon detector that can operate at temperatures below $-200^\circ$C,
a high-repetition-frequency short-pulse solid-state laser,
and a laser-rangefinder optical platform.
In addition to the reflection signal from the corner-cube retro-reflector on the relay satellite, the station  successfully received laser ranging signals echoed from all five retro-reflectors on the Moon.
The Earth-Moon distance can be precisely measured through the time interval between the moment the telescope emits laser beams and receives them back from the reflectors on the Moon.
The work of the TianQin laser rangefinder station has made China the third country to succeed in measuring the Earth-Moon distance through echo signals from  the lunar surface.
The measurement  provides centimeter precision on the location of the TianQin satellites, which showcased the advantages of a geocentric orbit in satellite navigation.

The main objective in step 1 is to launch the TianQin-1 satellite and test the inertial sensing, micronewton propulsion, drag-free control and laser interferometry technologies with in-orbit experiments.
Another purpose of TianQin-1 is to test the temperature control technologies and center-of-mass measurements of the satellite.
Work for step 1 started in 2016 and TianQin-1 recieved official approval from the China National Space Administration in 2018.
TianQin-1 was successfully launched on 20 December 2019 and all its mission goals have been achieved \cite{Luo:2020bls}.
The inertial sensor with a calibrated sensitivity of $5\times 10^{-12}\ \text{m\ s}^{-2}\text{Hz}^{-1/2}$ at 0.1 Hz
on the ground detected the residual acceleration of the satellite as about
$1\times 10^{-10}\ \text{m\ s}^{-2}\text{Hz}^{-1/2}$ at 0.1 Hz
and about $5\times 10^{-11}\ \text{m\ s}^{-2}\text{Hz}^{-1/2}$ at 0.05 Hz.
The thrust resolution is about 0.1 $\mu$N and the thrust noise is about
0.3 $\mu$N Hz$^{-1/2}$ at 0.1 Hz for the micronewton thrusters.
With drag-free control, the residual noise of the satellite is about
$3\times 10^{-9}\ \text{m\ s}^{-2}\text{Hz}^{-1/2}$ at 0.1 Hz.
The noise of the optical readout is about 30 pm Hz$^{-1/2}$ at 0.1 Hz.
The temperature control is about $\pm 3$ mK per orbit.
The mismatch in the center-of-mass measurements between the satellite and the test mass is about 0.1 mm.

For step 2, TianQin is scheduled to launch a pair of satellites (TianQin-2) in 2025 to make a major technological breakthrough in testing high-precision intersatellite laser interferometry.
In addition, TianQin-2 will use laser ranging to conduct high-precision distance measurements between the two satellites and  map the global gravity field.
TianQin-2 may detect GWs if it were not subject to substantial laser phase noise.
It is expected that the smooth advancement of step 2 will secure the final stage of TianQin project to deploy three satellites in geocentric orbit in 2035 and detect GWs.

TianQin project was jointly developed by Sun Yat-Sen University and Huazhong University of Science and Technology,
and more than 40 institutions in China have now joined the project.
International collaborations have always been encouraged and the international advisory committee was formed in 2018 to advise on the development of the TianQin project and invite international collaborations (\url{http://tianqin.sysu.edu.cn/en}).

\section{Taiji project}

In 2016, the Chinese Academy of Sciences announced the Taiji (太极) programme to detect GWs in space \cite{Hu:2017mde}.
In a heliocentric orbit, the external heat flux fluctuation of each satellite
can be minimized because the Sun-pointing angle of the satellite is very stable.
The center of mass of the constellation is about 1 au from the Sun
and the constellation has an inclination angle of 60$^\circ$ with respect to the ecliptic plane  to maintain the geometry of an equilateral triangle throughout the mission.
The normal vector of the detector plane rotates around the normal vector of the ecliptic plane, forming a cone with a $60^\circ$ half opening angle and a period of 1 yr.
Since the arm length of Taiji is longer than LISA,
it is more sensitive to low-frequency GWs emitted by SMBHs mergers, provided that Taiji and LISA have the same level of acceleration noise \cite{Guo:2018npi,tj2,tjptep}.
The design goal for the displacement noise is $\sqrt{S_x}=8\ \text{pm Hz}^{-1/2}$ and for the acceleration noise is
$\sqrt{S_a}=3\times 10^{-15}\
\text{m s}^{-2}\ \text{Hz}^{-1/2}$ at 1 mHz (ref. \cite{Ruan:2020smc}).
The noise curve is shown in Fig. \ref{noise}.

The primary target GW sources of Taiji are coalescing MBH binaries with total masses between $10^4$ to $10^8$ $M_\odot$.
Other sources include Galactic binaries,
the inspirals of stellar-mass BH binaries, BH-neutron star binaries,
the extreme- and intermediate-mass-ratio systems,
stochastic GW background and unknown sources.
Most Galactic binary sources constitute confusion noises for Taiji and LISA
because they cannot be resolved individually.
Expected key science motivators for Taiji are studies of galactic nuclei,
the formation history of SMBHs
and the evolution of stars in the Milky Way;
the role of mergers in galaxy evolution;
strong field environments close to MBHs; the nature of gravity;
and the physics of the early Universe.
Taiji will also address problems such as how the intermediate-mass seed BHs were formed,
how seed BHs grow
and whether dark matter could form a BH \cite{Hu:2017mde,Guo:2018npi,Ruan:2020smc,tjptep}.

According to the initial mission plan,
Taiji will be launched around 2033. A
roadmap towards Taiji’s final goal is given in three steps.
\begin{itemize}
  \item Step 1: develop and test Taiji’s
key technologies in ground laboratories.
To determine the shortfall between current capabilities and the final Taiji requirements,
a pre-pathfinder mission called Taiji-1 is planned to test the most important individual technologies,
including the laser interferometer, gravitational reference sensor, micronewton thruster, and drag-free control.
  \item Step 2: Taiji pathfinder (Taiji-2) is planned to be launched between 2023 and 2024.
Taiji-2 is a pair of technology demonstration satellites  designed to cover almost all of the Taiji technologies except the time-delay
interferometer.
  \item Step 3: After the technological obstacles are cleared by Taiji-2, the complete Taiji constellation (Taiji-3) is expected to launch in 2033  to detect GWs.
\end{itemize}

Taiji-1 was successfully launched on 31 August 2019, and all designed missions were completed \cite{tjptep}. The first-stage in-orbit test of Taiji-1 showed that
the accuracy of the displacement measurement of the laser interferometer reached the
order of 100 pm, the accuracy of the gravitational reference sensor on the satellite reached subnano-$g$
and the resolution of the microthruster was better than 1 $\mu$N.
The successful test meets the expectation of the Taiji-1 mission
design and shows that Taiji technology is feasible.

The Taiji pathfinder has a pair of satellites with a preliminary distance of $5\times 10^5$ km between them,
which will take the same orbit and have
the same satellite and payload designs as Taiji.
Individual spacecraft in the Taiji pathfinder will install only one laser interferometer system, forming a single-arm laser interferometer space antenna that could, in principle,  detect GWs if it were not subject to overwhelming laser phase noise.
One of the satellites will have two sets
of inertial sensors that will be arranged in the same configuration as Taiji.
Therefore, the drag-free control method with two
degrees of freedom can be tested.
To reduce the cost, there may be one inertial sensor
on the other spacecraft.

The laser for Taiji and Taiji pathfinder has a wavelength of 1064 nm and the prototype
has already been built in the Chinese Academy of Sciences.
A frequency stability of 30 Hz Hz$^{-1/2}$ of the laser has been
achieved.
The prototype of the Taiji interferometer has been built,
and the main functionalities have also been tested.
The precision of the displacement measurement achieves 15 pm Hz$^{-1/2}$,
1064 nm laser phase locking achieves 20 pm Hz$^{-1/2}$ and phasemeter readout precision reaches $2\pi\mu\text{rad Hz}^{-1/2}$.

To support the development of the Taiji programme with scientific, technical and management services,
the Gravitational Cosmic Taiji laboratory (Taiji Lab)
was established in the UNESCO International Center for Theoretical Physics Asia-Pacific (ICTP-AP).
The University of Chinese Academy of Sciences and other institutes
of the Chinese Academy of Sciences are involved in the Taiji programme.
The Taiji group has outlined the possibility of a direct collaboration with LISA (\url{https://ictp-ap.org/page/taiji-laboratory}).

\section{Future joint observations}
The planned space-based GW observatories LISA, TianQin and Taiji
will complement each other,
listen in the dark for any gravitational echo and provide an essential window into an unseen universe.
With the joint observations of these three detectors,
more sources and richer physics will be uncovered.
In addition to sky coverage, source localizations and parameter estimations,
joint space-based GW observations are promising for measuring GW polarizations and testing the theory of relativity.
The advantage of a two-detector network in space was first studied in ref. \cite{Crowder:2005nr}.

From Fig. \ref{noise}, we see that Taiji is similar to LISA except that it is $2-5$ times more sensitive than LISA above millihertz frequencies because of its longer arm length.
Although their abilities and scientific objectives are similar,
a combined network of LISA and Taiji could outperform individual detectors in localizing GW sources accurately and quickly.
It was shown that with the LISA-Taiji network,
the accuracy of the sky localization can be improved by
two orders of magnitude
for coalescing equal-mass BH binaries with a total mass $10^5 M_\odot$
at $z=1$ and $z=3$ (refs. \cite{Ruan:2019tje,Ruan:2020smc}).
The LISA-Taiji network improves
the angular resolution for various time-delay interferometry channels
by more than 10 times over each individual LISA or Taiji detector
for SMBH binaries with a mass ratio $q=1/3$ and masses $M_1=10^7 M_\odot$, $10^6 M_\odot$ and $10^5 M_\odot$ respectively for a primary BH
at $z=2$,
whereas the improvement from the LISA-Taiji network is moderate
for monochromatic GWs at 3 mHz and 10 mHz (ref. \cite{Wang:2020vkg}).
The search for polarization contents and parity violation
in the Stochastic GW Background with the LISA-Taiji network was discussed in
refs. \cite{Omiya:2020fvw,Orlando:2020oko,Wang:2021mou}.
It was shown that after 10 years of observation of the Stochastic GW Background around 1-10 mHz with the LISA-Taiji network,
the detection limit on extra polarizations other than the plus and cross modes could be improved by five orders of magnitude over the current upper bound around 10-100 Hz obtained with ground-based detectors \cite{Omiya:2020fvw}.
Employing a class of parameterized post-Einsteinian waveform \cite{Yunes:2009ke} with six parameters,
the measurements on amplitudes of alternative polarizations from joint LISA-Taiji observations could be improved by more than 10 times
compared with the LISA  mission only in an optimal scenario \cite{Wang:2021mou}.

TianQin is more sensitive to higher frequencies,
so it avoids the confusion noise and it is better suited to combine
with a third-generation ground-based detector for multi-band astronomy.
Considering that the LISA and TianQin detectors have different concepts and orbits,
it is expected that
they would be complementary to each other and that the LISA-TianQin network could improve the sky localization of GW sources.
For the Galactic double white dwarf binaries,
it was shown that the localization can be improved  up to three orders of magnitude \cite{Huang:2020rjf} compared with the single TianQin detector.
For monochromatic sources, it was found that TianQin performs better than LISA and Taiji at higher frequencies (above tens of millihertz) \cite{Zhang:2020hyx,Zhang:2020drf}.
The LISA-TianQin network can localize sources with frequencies in the range 1-100 mHz much more efficiently
and the network has greater sky coverage in terms of the angular resolution than  individual detectors \cite{Zhang:2020hyx,Zhang:2020drf}.

\begin{figure*}[htbp]
	\centering
	\includegraphics[width=0.9\textwidth]{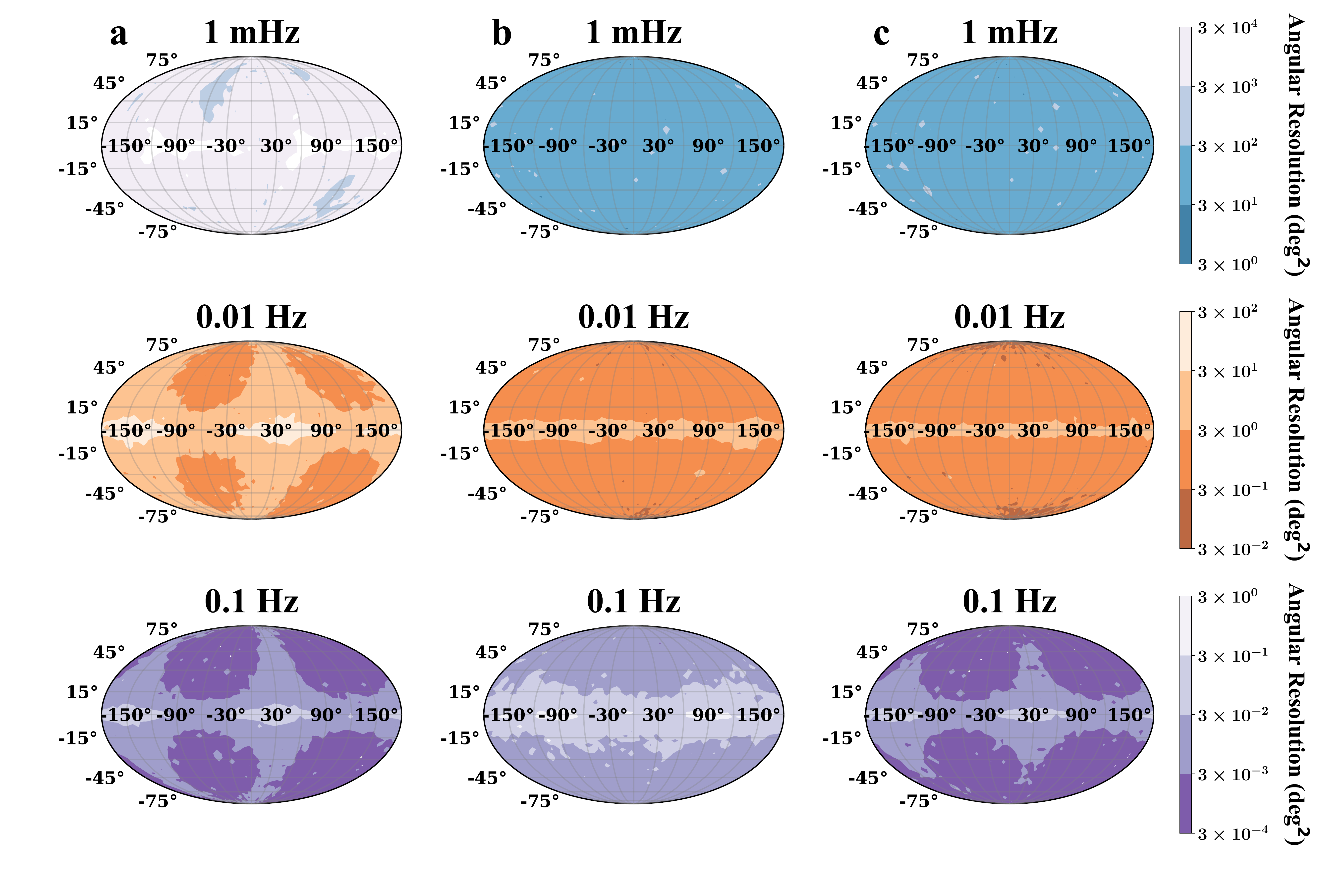}
	\caption{{\bf Sky maps for monochromatic sources.} The unit for the angular resolution is deg$^2$.
{\bf a,} Sky maps of angular resolution for TianQin at frequencies of 1 mHz (top), 10 mHz (middle) and 100 mHz (bottom).
{\bf b,} The corresponding sky maps of angular resolution for Taiji.
{\bf c,} The corresponding sky maps of angular resolution for the TianQin-Taiji network.}
\label{net1}
\end{figure*}

To explore the complementarity of the TianQin-Taiji network,
we took monochromatic sources with frequencies $10^{-3}$ Hz, $10^{-2}$ Hz and $10^{-1}$ Hz, respectively, and simulated
3,600 sources uniformly distributed in the sky with $-\pi/2<\theta_s<\pi/2$
and $-\pi<\phi_s<\pi$.
The SNR of the sources is chosen to be 7 with respect to LISA.
Since the Fisher information matrix approximation provides a good estimation of the sky localization of the source for a multi-detector network \cite{Shuman:2021ruh},
we used the Fisher information matrix method to estimate the angular resolutions for space-based GW detectors.
The mean angular resolution results  are shown in Table \ref{meantable} and the sky maps of the angular resolutions are illustrated in  Fig. \ref{net1}.
From the sky maps,
we see that at 1 mHz,
the angular resolution and sky coverage of Taiji are much better.
The reason is that at frequencies below several millihertz,
the contribution of the amplitude modulation
due to the annual rotation of the detector plane is similar to,
or even better than, the Doppler modulation,
and the amplitude modulation does not depend on angular locations of the sources \cite{Zhang:2020hyx}.
Above several millihertz, the amplitude modulation is negligible
and the Doppler modulation is dominant,
so the accuracy of the sky localization increases with the frequency and the sky maps have an equatorial pattern.
At 0.1 Hz, TianQin is more sensitive and performs better than Taiji.
As shown in Table \ref{meantable} and Fig. \ref{net1},
the TianQin-Taiji network improves the accuracy of the sky localization and increases the sky coverage greatly over individual detectors.
With the addition of LISA, the TianQin-Taiji-LISA network improves the accuracy of the sky localization by two times at 1 mHz,
whereas the improvements at 0.01 Hz and 0.1 Hz are small.

\begin{table}[htp]
  \centering
  	\begin{tabular}{|c|c|c|c|c|c|}
 \hline
 $f$/Hz & LISA & Taiji & TianQin & 2 Network & 3 Network\\ \hline
$10^{-3}$ & $2.0\times 10^{2}$ & $1.4\times 10^{2}$ & $1.2\times 10^{4}$ & $1.3\times 10^{2}$ & $69$ \\ \hline
$10^{-2}$ & $ 6.6 $ & $1.2$ & $8.2$ & $1.0$ & $0.79$ \\ \hline
$10^{-1}$ & $7.6\times 10^{-2}$ & $3.9\times 10^{-2}$ & $6.9\times 10^{-3}$ & $5.6\times 10^{-3}$ & $4.9\times 10^{-3}$ \\ \hline
\end{tabular}
\caption{{\bf The angular resolutions for monochromatic sources}.
The mean angular resolutions (in the unit
of deg$^2$) for monochromatic sources. 2 Network refers to the TianQin-Taiji network and 3 Network indicates the TianQin-Taiji-LISA network.}
    \label{meantable}
\end{table}

\begin{figure*}[htbp]
	\centering
	\includegraphics[width=0.9\textwidth]{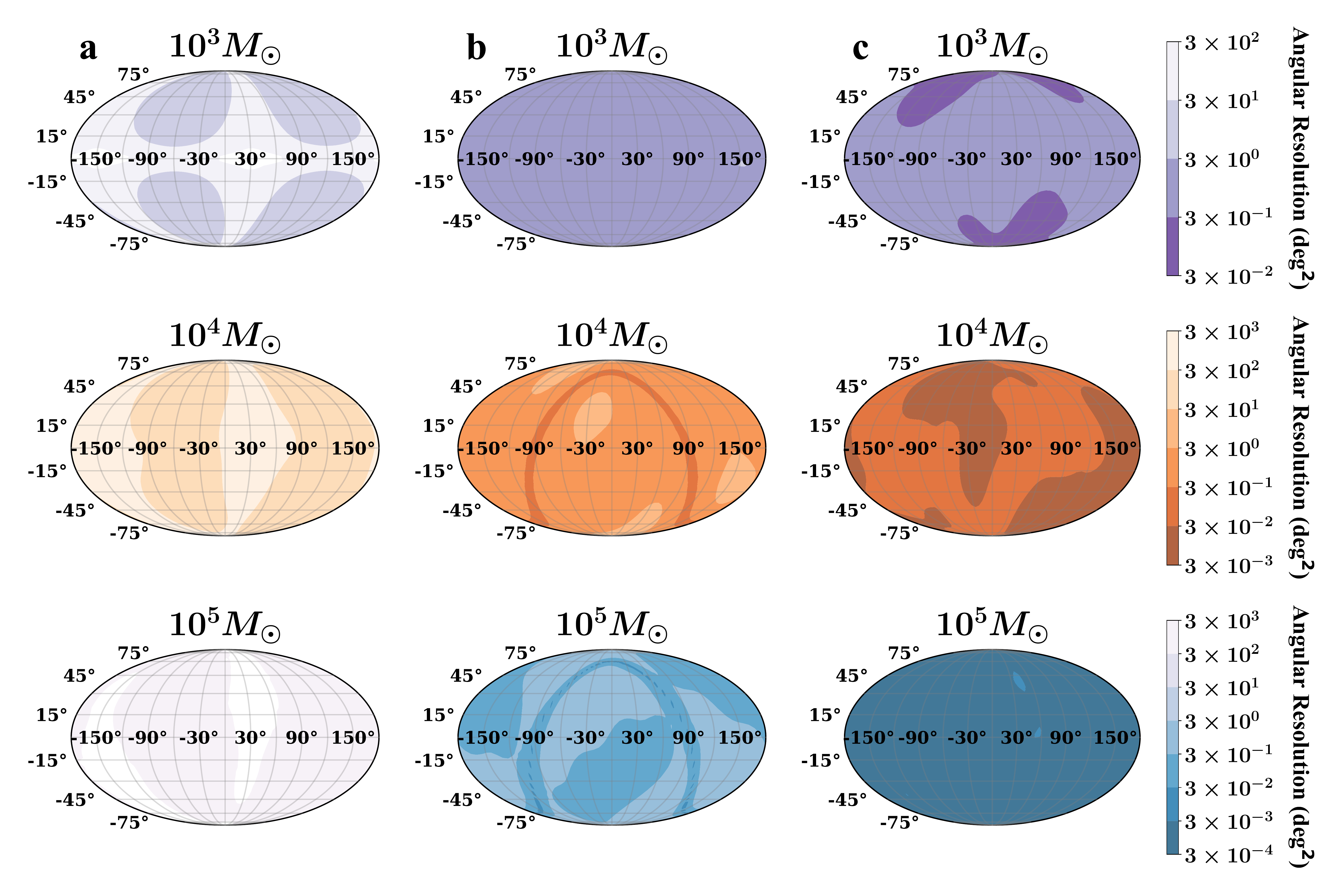}
	\caption{{\bf Sky maps for coalescence sources}.
 The unit for angular resolution is deg$^2$.
{\bf a,} Sky maps of angular resolution for TianQin for coalescence BBHs with equal masses of
$10^3 M_\odot$ (top), $10^4 M_\odot$ (middle) and $10^5 M_\odot$ (bottom) at $z=1$.
{\bf b,} The corresponding sky maps of angular resolution for Taiji.
{\bf c,} The corresponding sky maps of angular resolution for the TianQin-Taiji network.}
\label{net2}
\end{figure*}

We also simulate the coalescence of BBH with equal mass at $z=1$.
We take the component masses $m_1=m_2=10^3\ M_\odot$, $10^4\ M_\odot$ and $10^5\ M_\odot$,
and simulate the GW signals 1 yr before the coalescence.
The results are shown in Table \ref{meantable1} and Fig. \ref{net2}.
It is clear that the network of Taiji and TianQin improves the sky localization by almost two orders of magnitude for BBHs with the masses $m_1=m_2=10^4\ M_\odot$,
and more than two orders of magnitude for BBHs with the masses $m_1=m_2=10^5\ M_\odot$,
compared with the sky localization of Taiji alone.
The improvement using the network of LISA, Taiji and TianQin is a few times higher than the network of only Taiji and TianQin.
From Figs. \ref{net1} and \ref{net2},
we see that the high-angular-resolution regions of the sky are larger for the network,
so the network also improves the sky coverage.

\begin{table}[htp]
\begin{tabular}{|c|c|c|c|c|c|}
 \hline
 Mass($M_\odot$) & LISA & Taiji & TianQin & 2 Network & 3 Network \\
 \hline
 $10^3$ & 5.95 & 2.41 & $49.0$ & $0.72$ & $0.33$ \\
 \hline
 $10^4$ & 7.99 & 2.97 & $4.29\times10^{2}$  & $4.62\times10^{-2}$ & $1.65\times10^{-2}$ \\
 \hline
 $10^5$ & 2.42 & $0.64$ & $2.39\times10^{3}$ & $2.45\times10^{-3}$ & $6.19\times10^{-4}$ \\
 \hline
 \end{tabular}
\caption{{\bf The angular resolutions for coalescence sources}.
The mean angular resolutions (in the unit
of deg$^2$) for coalescence BBHs with equal mass at $z=1$.}
    \label{meantable1}
\end{table}

\section{Discussions}

GWs are truly a new window into the universe.
There are limits to what ground-based GW detectors can do--
we msut go into space if we want to push deeper into this new domain of understanding of the Universe.
In this Perspective we have introduced the concepts and status of two Chinese space GW detection projects, which will join LISA to deploy in the 2030s.
These two Chinese companions in the detection of GWs in space are not competitors,
but instead complement each other in listening to the symphony of GWs,
providing essential information about the unseen Universe.
While the goals of these Chinese projects are ambitious,
they require careful awareness of the difficulties of a single country going it alone on such a large project,
not only in terms of the cost,
but also in terms of science and technology resources.
Future possible collaborations between Chinese researchers and international experts are necessary.
International collaborations can open up a wide GW spectrum with a network of space-based facilities operated by China and other international players to improve the sky localization and sky coverage of GW sources,
examine more clearly the history of cosmic expansion and pull back the curtain on the mysteries surrounding the beginning of the Universe.
Such collaborations could provide information on
the formation of SMBHs and strong field environments close to MBHs,
and allow us to better understand no hair theorem, polarization contents and the nature of gravity.

\end{CJK*}

\textbf{Data availability}

The data that support the findings of this study are available from the corresponding
author Y.G. upon reasonable request.
The data for Fig. 2 can be generated from the code deposited in \url{https://github.com/yggong/transfer_function}.

\textbf{Code availability}

The python code can be obtained at \url{https://github.com/yggong/transfer_function}.

\textbf{Acknowledgements}
This research was supported by the National Key Research and Development Program of China under grant numbers 2020YFC2201504 and 2020YFC2201400,
the National Natural Science
Foundation of China under grant numbers 11875136 and 12075202, and
the Major Program of the National Natural Science Foundation of China under grant number 11690021.
B.W. acknowledges the support from Shanghai Education Commission.
J.L. acknowledges the support from Guangdong Major Project of Basic and Applied Basic Research under grant number 2019B030302001.

\textbf{Author contributions}
All authors contributed to the work presented in this paper.
Y.G. analysed the data, contributed analysis tools and wrote the paper.
J.L. conceived TianQin and reviewed the paper.
B.W. contributed materials and wrote the paper.

\textbf{Competing interests}
The authors declare no competing interests.

\textbf{Additional information}
Correspondence should be addressed to Yungui Gong or Bin Wang.

\end{document}